\documentclass[
 reprint,
superscriptaddress,
 amsmath,amssymb,
 aps,pra
]{revtex4-2}

\usepackage{graphicx}
\usepackage{dcolumn}
\usepackage{bm}
\usepackage[colorlinks=true,linkcolor=blue,urlcolor=blue,citecolor=blue]{hyperref}

\usepackage{subfigure}
\usepackage{xcolor}
\usepackage{amsmath}
\usepackage{physics}
\usepackage{float}
\begin{document}

\preprint{APS/123-QED}
\title{Probing inhomogeneous and dual asymmetric angular momentum exploiting spin-orbit interaction in tightly focused vector beams in optical tweezers}

 \author{Ram Nandan Kumar}
\email{ramnandan899@gmail.com}
 \affiliation{Department of Physical Sciences, Indian Institute of Science Education and Research Kolkata, Mohanpur-741246, West Bengal, India}

 \author{Jeeban Kumar Nayak}
 \affiliation{Department of Physical Sciences, Indian Institute of Science Education and Research Kolkata, Mohanpur-741246, West Bengal, India}

 \author{Anand Dev Ranjan}
 \affiliation{Department of Physical Sciences, Indian Institute of Science Education and Research Kolkata, Mohanpur-741246, West Bengal, India}

\author{Subhasish Dutta Gupta}
\affiliation{Department of Physical Sciences, Indian Institute of Science Education and Research Kolkata, Mohanpur-741246, West Bengal, India}
\affiliation{School of Physics, Hyderabad Central University, India}
\affiliation{Tata Institute of Fundamental Research Hyderabad, India}

\author{Nirmalya Ghosh}
\email{nghosh@iiserkol.ac.in}
\affiliation{Department of Physical Sciences, Indian Institute of Science Education and Research Kolkata, Mohanpur-741246, West Bengal, India}

\author{Ayan Banerjee}
 \email{ayan@iiserkol.ac.in}
 \affiliation{Department of Physical Sciences, Indian Institute of Science Education and Research Kolkata, Mohanpur-741246, West Bengal, India}

\date{\today}
\date{\today}

\begin{abstract}

The spin-orbit interaction (SOI) of light generated by tight focusing in optical tweezers has been regularly employed in generating angular momentum - both spin and orbital - in trapped mesoscopic particles. Specifically, the transverse spin angular momentum (TSAM), which arises due to the longitudinal component of the electromagnetic field generated by tight focusing, is of special interest, both in terms of fundamental studies and associated applications. We provide an effective and optimal strategy for generating TSAM in optical tweezers by tightly focusing radially and azimuthally polarized first order Laguerre Gaussian beams with no intrinsic angular momentum, into a refractive index stratified medium. Our choice of such input fields ensures that the longitudinal spin angular momentum (LSAM) arising from the electric (magnetic) field for the radial (azimuthal) component is zero, which leads to the separate and exclusive effects of the electric and magnetic TSAM  in the case of input radially and azimuthally polarized beams on single birefringent particles. We also observe the emergence of origin-dependent intrinsic orbital angular momentum causing the rotation of birefringent particles around the beam axis for both input beam types, which opens up new and simple avenues for exotic and complex particle manipulation in optical tweezers.
\end{abstract}


\maketitle


\section{Introduction}

The effects of spin-orbit interaction (SOI) - which couples the spin and orbital degrees of freedom of light - have been particularly useful in inducing intriguing rotational dynamics in particles confined by optical tweezers, including both spin motion around the particle axis and orbital motion around the beam axis \cite{yang2021optical,kotlyar2020spin,zhao2007spin}. The spin motion can be induced using tightly focused spin-polarized (left or right circular polarization) Gaussian beams which exchange their longitudinal spin angular momentum (LSAM) with birefringent micro-particles \cite{friese1998optical,garces2003observation}, while orbital motion is typically induced using beams that carry intrinsic orbital angular momentum (OAM)\cite{stilgoe2022controlled,he1995direct,anhauser2012acoustic}. Another comparatively exotic spin motion arises due to the transverse spin angular momentum (TSAM) - which arises as a direct consequence of the longitudinal component of the electric (or magnetic) field which is often at the heart of SOI of light \cite{novotny2001longitudinal,shao2018spin,roy2022manipulating}. While TSAM has been studied in detail in theory \cite{bliokh2015transverse,bliokh2013dual,bekshaev2015transverse}, experimental evidence has been obtained mostly in the case of evanescent fields, in the form of rotation of dielectric particles  \cite{aiello2015transverse, tkachenko2020light,bliokh2014extraordinary}.  However, obtaining signatures of TSAM for propagating fields are difficult to obtain experimentally since these are often conjugated with those due to LSAM - leading to complex rotational motion in the probe particles \cite{li2017transverse}. 

To address this issue, a strategy using co-propagating opposite circularly polarized fundamental Gaussian beams was devised recently \cite{stilgoe2022controlled} in optical tweezers, where the opposite nature of the helicity ensured that the LSAM cancelled out, leading to only TSAM being present near the focal plane - the effects of which were observed on trapped birefringent particles. However, this is not a direct method of generating TSAM, and the challenge is thus to produce beams which lead to clear and unambiguous effects of TSAM on probe particles.  For this purpose, an interesting candidate may be radially and azimuthally polarized $m=0$ LG  beams which have zero intrinsic OAM, but possess an intensity zero on the beam axis \cite{zhan2009cylindrical}. Tight focusing of such beams lead to the generation of a significant longitudinal field component \cite{dorn2003sharper,sato2009hollow,liu2019separation,huang2011vector}, while the absence of an intrinsic OAM could produce intriguing effects. On another note, instances are well known in wave optics where the so-called 'electromagnetic democracy' \cite{berry2009optical} breaks down due to the electromagnetic asymmetry of matter (as is the case in metals, Mie scattering, etc.). Indeed, the symmetry of the electric and magnetic fields in the context of the angular momentum (AM) of light has been studied theoretically \cite{bliokh2013dual}, and dual-asymmetric TSAM has also been discussed earlier \cite{bliokh2015transverse}.  It is still tempting to ask the question: Can the effects of the electric and magnetic fields be separately determined experimentally in the context of the AM of light? 

In this paper, we attempt to answer this interesting question. We tightly focus radially and azimuthally polarized LG beams of $m=0$ in an optical tweezers setup and observe the consequences of tight focusing on birefringent microparticles. Additionally, the tightly focused beams also propagate through a refractive index (RI) stratified medium before they are incident into the trapping region inside our sample chamber. The influence of the stratified medium is crucial in determining the interaction of the light beam with the particles and influencing their dynamics as we had observed earlier \cite{roy2013controlled,roy2014manifestations,pal2020direct}. Here, the tightly focused radially or azimuthally polarized light passing through a stratified medium induces different spin dynamics in birefringent particles depending on their spatial location in the trapping region. The longitudinal field that is generated due to tight focusing of the input LG beam by the high numerical aperture (NA) objective lens, which is integral to optical tweezers, gives rise to a finite intensity at the beam center \cite{dorn2003sharper,youngworth2000focusing}. This also ensures that there is a finite TSAM at focal region, while - most importantly - the LSAM is zero by construction. Thus, any rotational motion seen for spherical birefringent particles will be exclusively due to the TSAM. Indeed, a highly birefringent liquid crystal droplet appears spinning at the trap center as observed using polarization-based imaging - a clear manifestation of TSAM. Most importantly, radially and azimuthally polarized light gives rise to {\it purely} electric and magnetic TSAM, respectively, which are expressed separately on the liquid crystal particles. Hence, it appears that the symmetry of the electromagnetic field is broken in this instance, and the effects of the electric and magnetic field can be separately observed experimentally due to the choice of our structured beam. In addition, we observe orbital motion in particles trapped in the annular intensity ring around the trap centre due to the intrinsic OAM, which is developed with respect to the centre of gravity (axis) of the beam ($r\times p$, where $p$ is the total canonical momentum) \cite{o2002intrinsic}. We carry out rigorous numerical simulations of our system for different RI values of the stratified medium that the tightly focused light encounters as it propagates \cite{kumar2022probing}, which helps us choose the most appropriate value of the RI contrast of the stratified medium to obtain the best experimental results.  In what follows, we describe the basic theoretical premise of our work.

\section{Theory}
We now determine analytical expressions for the spin angular momentum (SAM) and total OAM densities for tightly focused radially and azimuthally polarized LG ($m=0$) beams, and show that the LSAM is zero, while the contribution for TSAM in the case of input radially polarized light comes from the electric fields, while that for input azimuthally polarized light comes from the magnetic fields, solely. For this, we first note that  a tightly focused radially polarized LG ($m=0$) beam contains all three components of electric fields ($E_{x}$, $E_{y}$ and $E_{z}$), and the transverse magnetic field components ($H_{x}$ and $H_{y}$), $H_{z}$ being 0. However, an azimuthally polarized $LG_{10}$ beam contains all three componets of magnetic field ($H_{x}$, $H_{y}$ and $H_{z}$), and the transverse  electric field components ($E_{x}$ and $E_{y}$), $E_{z}$ being 0.  Now, the time-averaged Poynting vector $\mathbf{P}$ for a monochromatic electromagnetic field is $\bold{\mathbf{p}=\varepsilon_0\langle\mathbf{E} \times \mathbf{B}\rangle}
$, while the total OAM density is $\bold{L=r \times P}$ \cite{o2002intrinsic}. This is an origin-dependent quantity and depends upon the lateral position of the corresponding axis \cite{o2002intrinsic}. On the contrary, SAM ($\mathbf{S}$) is intrinsic in nature (origin independent). Thus, $\mathbf{S} \propto \operatorname{Im}\left[\epsilon_0\left(\mathbf{E}^* \times \mathbf{E}\right)+\mu_0\left(\mathbf{H}^* \times \mathbf{H}\right)\right]$ or   ($\mathbf{S}=\mathbf{S}^{\mathbf{e}}+\mathbf{S}^{\mathrm{m}}$) with  $\epsilon$ being the permittivity,  $\mu$ the permeability, $\mathbf{S}^e$  and $\mathbf{S}^m$ are electric and magnetic spin angular momentum density of light respectively \cite{neugebauer2015measuring}. Hence, the SAM and total OAM density for the radially (azimuthally) polarized LG ($m=0$) beams on tight focusing in optical tweezers maybe written as
\begin{eqnarray}
&S_{x}=\operatorname{Im}\{\left(-C i\left(I_{11} I_{10}^{*}+I_{11}^{*} I_{10}\right) \sin \phi\right)\} \nonumber\\
&S_{y}=\operatorname{Im}\{\left (C i\left(I_{11} I_{10}^{*}+I_{11}^{*} I_{10}\right) \cos \phi\right )\}\nonumber\\
&S_{z}=0
\end{eqnarray}
\begin{eqnarray}
&L_{x}=\operatorname{Re}\left\{D\left(-y I_{11} I_{12}^{*}-i z I_{12}^{*} I_{10} \sin \phi \right)\right\}.\nonumber \\
&L_{y}=\operatorname{Re}\left\{D\left(x I_{11} I_{12}^{*}+i z I_{12}^{*} I_{10} \cos \phi\right)\right\}. \nonumber\\
&L_{z}=\operatorname{Re}\left\{D\left(i x I_{12}^{*} I_{10} \sin \phi-i y I_{12}^{*} I_{10} \cos \phi\right)\right\}.
\end{eqnarray}
where, C and D are constants corresponding to SAM and OAM of a radially (azimuthally) polarized 1st order LG $m=0$ beam.
$I_{10}$, $I_{11}$ and $I_{12}$ are the Debye-Wolf integrals \cite{roy2013controlled}; and $\phi$
is the azimuthal angle in the cylindrical (or spherical) coordinate
system. 

 Now, since $H_z$ is zero for radially polarized light, while $E_z$ is zero for azimuthally polarized light, the contributions to TSAM are only from the electric field for radially polarized light ($\mathbf{S}_{\perp}^{\mathrm{e}} \neq 0, \quad \mathbf{S}_{\mathrm{z}}^{\mathrm{e}}$ and $\mathbf{S}^{\mathrm{m}}=0$), and from the magnetic field for azimuthally polarized light ($\mathbf{S}_{\perp}^{\mathrm{m}} \neq 0, \quad \mathbf{S}_{\mathrm{z}}^{\mathrm{m}}$ and $\mathbf{S}^{\mathrm{e}}=0$).   In addition, besides being of separate independent origin, the TSAM is also rather large due to the tight focusing \cite{pal2020direct}, and is capable of causing transverse spin of a birefringent liquid crystal droplet about its own axis. Interestingly, the effects of magnetic TSAM have been generally neglected in the literature, where the focus of interest has typically been the electric component. On another note, the total OAM can cause the rotation of birefringent particles around the beam propagation ($z$) axis.  We now proceed to numerically simulate our experimental system next, and determine the TSAM and total OAM characteristics.
 
\onecolumngrid
\begin{center}
\begin{figure}
\includegraphics[width=\textwidth]{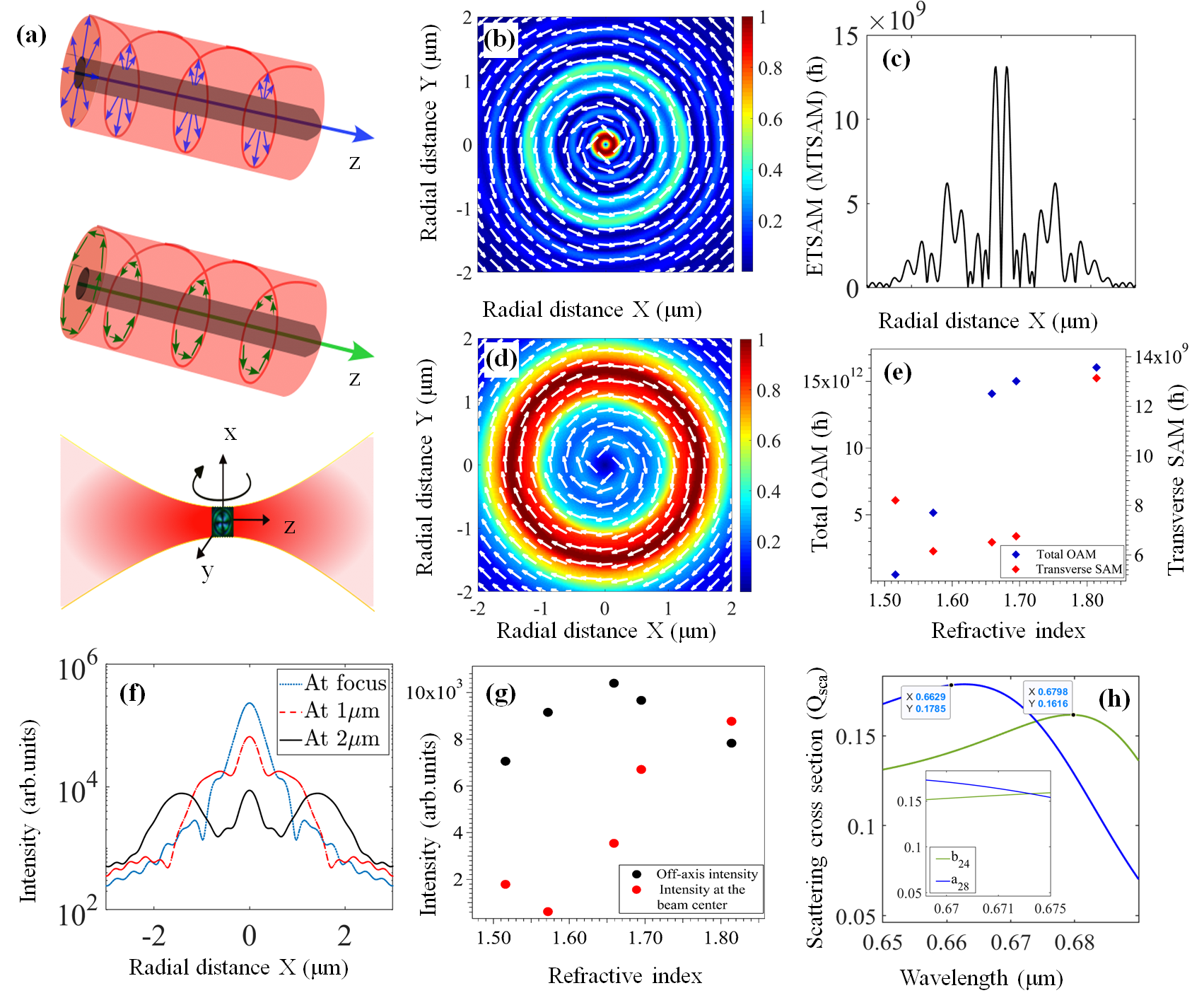}
\caption{(a) Schematic diagrams of radially and azimuthally polarized structured vector beams with an illustration of birefringent LC droplet undergoing transverse spinning about its axis. (b) Simulation of TSAM at $2~ \mu$m way from the focus of radially (azimuthally) polarized structured vector beams for mismatched RI (1.814). (c) Radial distribution of electric (magnetic) TSAM at $2~ \mu$m way from the focus which is highly concentrated at the beam center. (d) Simulation of total OAM at $2~ \mu$m way from the focus of radially (azimuthally) polarized structured vector beams for mismatched RI (1.814). (e) Total OAM and transverse SAM as a function of RI. (f) Line plot of the intensity distribution of a radially polarized beam at different $z$-planes for mismatched RI (1.814). (g) Comparison of intensities at the trap center (red solid circles) and off-axis (black solid circles) as a function of RI at 2$\mu$m away from the focus. (h) Comparison of electric and magnetic type scattering modes ($a_{n}$ and $b_{n}$, respectively) excited in the trapped LC by radially polarized light, with the inset showing the mode characteristics around the trapping wavelength 0f 671 nm }
\label{OAM}
\end{figure}
\end{center}

\twocolumngrid
\section{ Numerical simulations} 
In our experimental system, the output from a high NA objective lenses in optical tweezers setup is passed through a stratified medium. The laser beam of wavelength 671 nm is incident on the 100X oil immersion objective of NA 1.4 followed by (a) an oil layer of thickness around 5 $\mu m$ and refractive index (RI) 1.516, (b) a 160 $\mu m$ thick coverslip having refractive index varying between 1.516-1.814 (note that the case where the $RI = 1.516$ is henceforth referred to as the ”matched condition,” which is typically employed in optical tweezers to minimize spherical aberration effects in the focused beam spot, whereas the other values are referred to as a 'mismatched' condition) (c) a sample chamber of an aqueous solution of birefringent RM257 particles and liquid crystal droplets in a water medium having a refractive index of 1.33 with a depth of 35 $\mu m$, and finally (d) a glass slide of refractive index 1.516 whose thickness we consider to be semi-infinite ( 1500 $\mu m$) [see Fig. \ref{schematic} (I)]. In the simulation, the origin of coordinates is taken inside the sample chamber at an axial distance of 5 $\mu m$ from the interface between the sample and the coverslip. Thus, the objective-oil interface is at -170 $\mu m$, the oil-cover slip interface is at -165 $\mu m$, and the cover slip-sample chamber interface is at -5 $\mu m$, and the sample chamber-glass slide interface is at +30 $\mu m$. 

Fig. ~\ref{OAM} (a) is a cartoon representation of our system, while the results of our simulations are shown in Fig. ~\ref{OAM} (b)-(h). For the TSAM (Fig. ~\ref{OAM} (b) and (c)) and OAM Fig. ~\ref{OAM} (d), we show results for the mismatched RI - since both quantities are highest for an RI of 1.814, which we show in Fig. ~\ref{OAM} (e). Also, the spherically aberrated intensity profile that we obtain in this case allows an overlap between large intensity and large TSAM/OAM that is useful to see effects on mesoscopic particles of diameter a few microns \cite{pal2020direct}. Note that we also perform the simulations not at the focal region of the trap, but at 2 $\mu m$ away from the focus - so as to obtain enough spatial extent of both intensity and TSAM/OAM to obtain experimentally discernible effects. The corresponding intensity distributions as a function of axial distance from focus for the mismatched case, and a comparison of intensities at the beam center and off-axis as a function of RI are shown in Fig. ~\ref{OAM} (f) and (g), respectively.
\onecolumngrid
\begin{center}
\begin{figure}[!h]
\includegraphics[width=\textwidth]{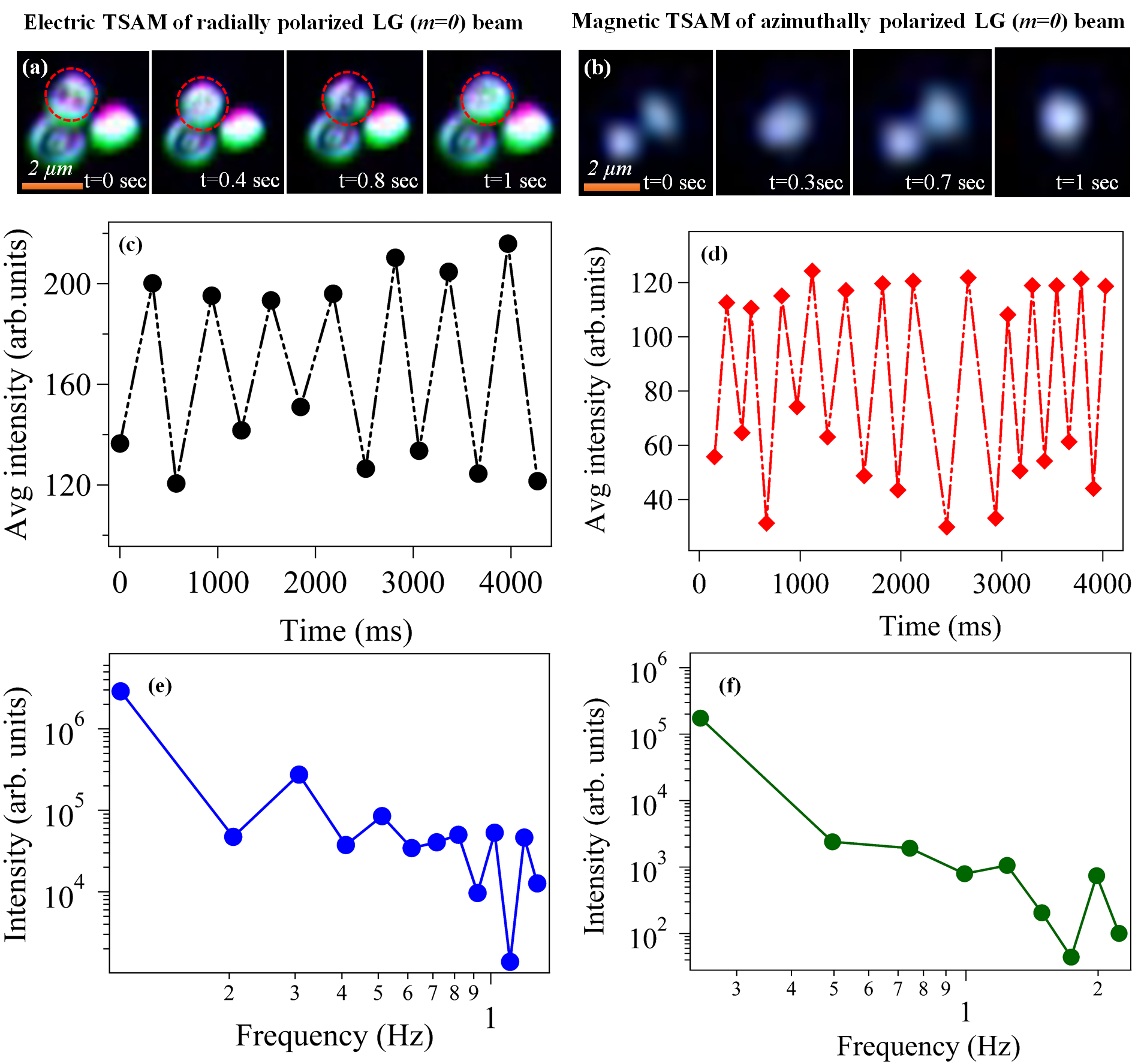}
\caption{ Experimental results:  (a) and (b) show time-lapsed frames of videos (Videos1 and 2, respectively, provided in the online Supplementary Information) in the cross-polarization scheme showing the orientation of lobes of an LC particle on tight focusing of radially (a) and azimuthally (b) polarized LG $m=0$ beams, respectively. (c) and (d) show the frequency of spinning of the LC particle about its transverse axis (xy) due to electric and magnetic TSAM of radially (c), and azimuthally polarized (d) LG $m=0$ beams, respectively. Note that the liquid crystal (LC) droplet in the red circle indicates the one that probes the effect of transverse SAM, while the other particles are merely trapped due to the intensity gradient of light. (e) and (f) The highest frequency component of spinning of the LC particle is around 1 Hz for radially polarized light and around 2 Hz for azimuthally polarized light respectively.}
\label{expresults}
\end{figure}
\end{center}
\twocolumngrid
The quiver plot in Fig.~\ref{OAM}(b) demonstrates that the TSAM for the mismatched RI case is in the anticlockwise direction. Note that this is the entity that leads to the spin of a LC particle about its transverse axis lying in the xy-plane, as shown in the cartoon in Fig.~\ref{OAM}(a). Fig.~\ref{OAM} (c) demonstrates the distributions of electric and magnetic TSAM - from which it is clear that the TSAM peaks at the centre which suggests that a LC particle trapped at the focus would feel its effects the most. Fig.~\ref{OAM}(d) demonstrates the quiver plot of the total OAM, which is also in the clockwise direction - and is in an annular ring around the focus. We envisage that particles trapped in this ring would exhibit rotational motion around the beam centre. In this connection, Fig.~\ref{OAM}(f) shows that the spatial extent of the radial intensity distribution for the input radially polarized beam is understandably highest at 2 $\mu m$ away from the focus - and we observe from experiments, that the intensity gradient is also enough to trap particles in this region.

We then focus on the possibility of separately probing the effects of electric and magnetic TSAM for input radial and azimuthal polarizations, respectively, on our LC particles. The efficacy of coupling of electric/magnetic fields of light with the scattering particles depends on whether the corresponding electric $a_{n}$ (transverse magnetic-TM) or magnetic $b_{n}$ (transverse electric-TE) modes are efficiently excited by the longitudinal electric/magnetic fields of the incident optical field. It is therefore possible to discern and separately observe the mechanical action of the electric and the magnetic TSAM of the optical field on a spherical probe particle by looking for the dominance of either modes of the probe particle at the wavelength of interest. Using a standard Mie scattering algorithm, and the RI profiles of our LC particles, we determine these in Fig.~\ref{OAM}(h) as a function of wavelength.  It is clear that both have peaks in the vicinity of 671 nm, but the peak of the dominating $b_{n}$ mode (identified to be n= 24) is closer to that wavelength compared to that of the corresponding  dominant $a_{n}$ mode (n = 28). Thus, we expect higher transfer of TSAM for the input azimuthally polarized light - where the TSAM is magnetic in nature - which would possibly lead to higher rotation frequency in this case.

\section{ Experimental results}
\begin{center}
\begin{figure}
\includegraphics[width=0.5\textwidth]{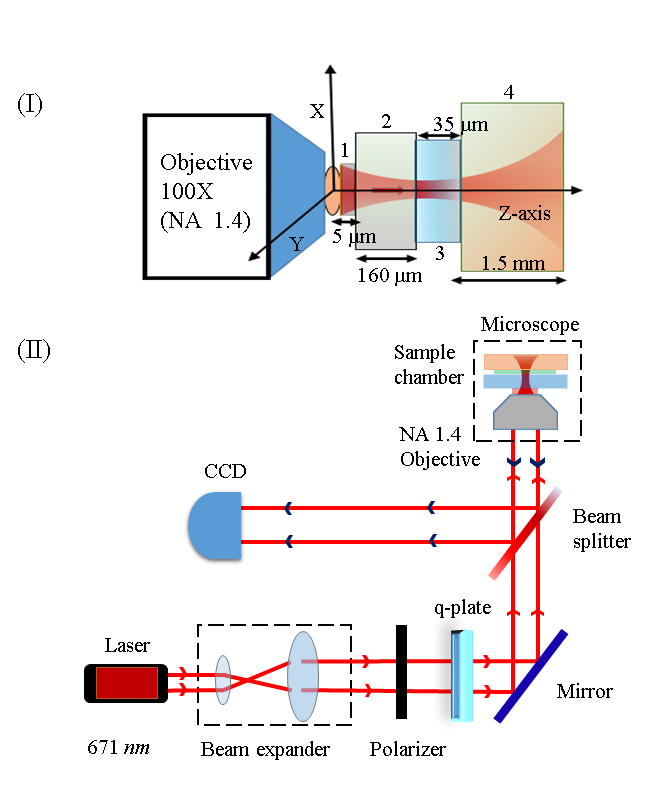}
\caption{Schematic diagram of our experimental set-up. (I) Illustration of the stratified medium used in our numerical
simulation as well as in the microscope. (II) Ray diagram of the tightly focused radially (azimuthally) polarized beams in optical tweezers.  }
\label{schematic}
\end{figure}
\end{center}

We now proceed to our experimental results which are shown in Fig.~\ref{expresults}(a)-(f). The schematic and details of our optical tweezers setup are provided in the experimental methods section (Fig.~\ref{schematic} (II)).  We use a vortex half-wave retarder ($q$-plate) of zero-order for generating structured vector beams (i.e.radially and azimuthally polarized LG $m=0$ beam), with input linear $x$- polarized and $y$-polarized light which are converted into azimuthally and radially polarized light, respectively. We use RM257 vaterite  and nematic liquid crystals as the probe particles - they being optically anisotropic and birefringent so as to transfer angular momentum (orbital and spin) from the beam into the particles. The mean size of RM257 particles is $ 1-2 ~\mu$m, while that of LC droplets are $ 2-4 ~\mu$m  with a standard deviation of 20\%. The LC droplets have much higher birefringence compared to the vaterite particles, so that we use them to probe effects of TSAM. The RM257 particles, on the other hand, are much smaller - so that they can be trapped in the annular intensity ring, in order to probe the effects of OAM. The transfer of TSAM to particles trapped at the trap center, as well as the transfer of OAM to particles trapped in the off-axis intensity ring are optimized by varying the $z$-focus of the microscope objective. 
\begin{center}
\begin{figure}[!h]
\includegraphics[width=0.5\textwidth]{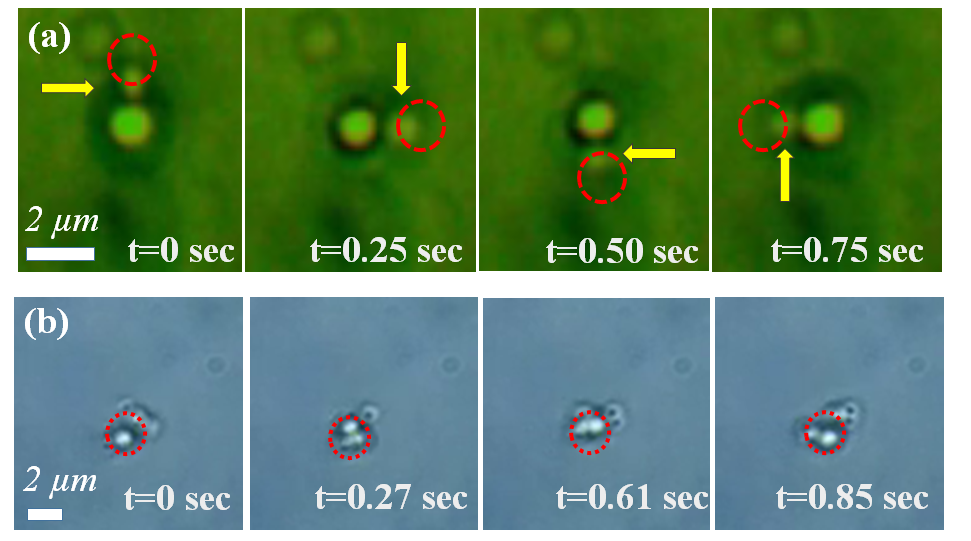}
\caption{ Time-lapsed frames of a video recording (Videos3 and 4 in online Supplementary Information) showing the rotation of particles by tightly focused radially (azimuthally) polarized LG $m=0$ beam. (a) The red circles mark the trajectory of an RM257 birefringent particle rotating around another trapped particle at the center of the beam, but at a different axial depth. (b) The red circle show the orbit of rotation of similar particles at 2$\mu$m away from the focus of azimuthally polarized LG $m=0$ beam. There is zero intensity ($E_{z}=0$) at the center of the beam so particles are only trapped or orbiting in an annular ring at different axial distances.
}
\label{exp_OAM}
\end{figure}
\end{center}
For discerning the effects of TSAM, we use a cross-polarization scheme - where use crossed linear polarizers at the input and output of the microscope. The rotation of the LC particles under the influence of radial and azimuthally polarized light is shown in Fig. ~\ref{expresults}(a) and (b) (the respective videos, Video1 and 2, are provided in the online Supplementary Information), respectively. Due to the cross-polarization, four polarization lobes appear across the surface of the LC particle in accordance with its scattering properties. These lobes clearly appear to be spinning, as is shown in their different spatial locations in Fig.~\ref{expresults}(a) and (b). Note that the use of crossed polarizers to discern transverse spin referred to as `pitch rotation' has been used earlier in Ref.~\cite{vaippully2021continuous}. The lobes also move in the laterally across the image - ascertaining the rotation to be indeed in the $x-z$ plane. The particle which is spinning is shown encircled in red. The other particles adjoining it do not sample TSAM, and are merely trapped by the intensity gradient of light. This indeed proves the spatially inhomogeneous distribution of TSAM that  our simulations predict. The rates of rotation are depicted in Figs.~\ref{exp_OAM}(c) and (d) from frame-by-frame analysis of the videos. The rotation is not very regular - which is understandable considering the variation of TSAM across the surface of the particle which possibly leads to fluctuations in the overall motion, and implies that the final signal may be a superposition of different rotational frequencies. 

To determine the rotational frequencies, we perform fourier transforms (sampling frequency around 5 Hz) of this signal extracted from the video - which are shown in Figs.~\ref{expresults}(e) and (f).  We apply a Hanning window function to efficiently extract the peak amplitudes as a function of frequency, and observe that while there appear multiple rotation peaks in both sets of data,  the dominant peak is higher in Fig.~\ref{expresults}(f) (around 2 Hz) compared to that in Fig.~\ref{expresults}(e) (several peaks between 0.3-1.2 Hz). This indicates that the rotation due to the azimuthally polarized light, which couples with the magnetic scattering modes, is faster compared to that due to the radially polarized light, since the former is more prominent than the electric scattering modes as we showed in Fig.~\ref{OAM}(h). These results also clearly indicate our ability to experimentally discern the effects of the electric and magnetic fields of light through the coupling of TSAM for radially and azimuthally polarized input light with the LC particles, respectively. 

Next, we consider the effect of OAM on the much smaller and less birefringent RM257 particles. With radially polarized light, we observe in Fig.~\ref{exp_OAM}(a) that a single large particle (size around 2 $\mu$m) is trapped at the beam center, while another smaller particle (size around 1 $\mu$m) is rotating in the annular ring in time lapsed images (see Video3 in the online Supplementary Information). Note that the particle in the center (which is also at a different axial distance with respect to the spinning particle) does not seem to spin - which we believe is due to the lower birefringence of vaterite compared to the LC. In some cases, we even obtain multiple particles trapped in the annular ring (not shown here), which also appear to move in the ring - but the movement does not appear to be synchronized. We are presently working to observe more systematic effects for multiple particles trapped in our configuration. In the case of azimuthally polarized light, RM257 particles are not trapped at the center of the beam because the longitudinal component of the electric field is zero, so we observe single particles as well as clusters, rotating around the beam center (see Video4 in the online Supplementary Information), as we show in the time lapsed images in Fig.~\ref{exp_OAM}(b). A few particles that do not appear to spin are possibly at different axial distances where the OAM is lower. Thus, from these experiments clearly demonstrate that tight focusing of radially and azimuthally polarized light generates OAM that helps in rotating the particles about the beam propagation axis, even though the beam does not contain any intrinsic OAM. The values of OAM for both radial and azimuthal polarized light are same in the region we consider, since it depends on the radius of the annular intensity distribution (origin dependent) and the longitudinal component of the electric (magnetic) field in that region. We also observe that the frequency of rotation increases as we increase the power. This is expected as the magnitude of both the electric and magnetic fields will increase on increasing the input intensity. 

\section{Conclusion}
In conclusion, we study the SOI of light generated due to the tight focusing of structured vector beams in optical tweezers to engineer the dynamics of birefringent micro-particles and liquid crystal at different spatial locations close to the focal region of the tweezers. Thus, we tightly focus radially and azimuthally polarized vector $m=0$ (Laguerre-Gaussian) beams - that do not carry any intrinsic orbital angular momentum (OAM) - into a refractive index stratified medium and observe the effects of both TSAM and OAM on single birefringent particles trapped in the trap center, and single or multiple birefringent particles orbiting around the beam propagation axis, respectively. Our configuration is rather unique in the sense that the LSAM for such vector beams is zero by construction, so that any rotation we observe about the particle body axis is purely due to TSAM - which is generated due to the longitudinal component of the field that arises due to tight focusing. Our system also allows us to probe the effects of electric and magnetic TSAM separately, which we show for input radially and azimuthally polarized beams, respectively. In addition, we see clear signatures of origin-dependent OAM generated for both input polarizations, that we are able to observe experimentally on birefringent particles due to the spherical aberrated intensity profile generated by our RI stratified medium. Thus, our work provides an experimentally viable strategy for engineering optical traps with controlled and specific, yet variable, spin-dynamics - including unambiguous signatures of TSAM - of trapped particles at different spatial regions near the trap focus. Importantly, this is due to SOI effects generated by tight focusing alone, without the need for structuring complex beam profiles using advanced algorithms involving adaptive optics. In the future, we plan to observe the effects of tight focusing and stratification on more complex structured beams, and even work on ENZ (Epsilon Near Zero) materials, to devise interesting routes of generating complex particle trajectories in optical tweezers.\\

\textbf{Acknowledgements} \par 
The authors acknowledge the SERB, Department of Science and Technology, Government of India (Project No. EMR/2017/001456) and IISER Kolkata IPh.D fellowship for research. They also acknowledge Sauvik Roy for their help in simulation.\\
\medskip

\textbf{Author Contributions}\par
R.N.K. and A.B. conceived the idea; R.N.K. performed the experiment, analyzed the data and did corresponding numerical simulations; J.K.N. performed the Mie theory based analysis; A.D.R. prepared the RM257 samples and helped R.N.K. to build the set-up; S.D.G., N.G. and A.B. supervised the overall project. R.N.K., A.B., N.G. and S.D.G. wrote the manuscript. All the authors discussed the results.\\
\medskip

\section{Appendix}
\subsection{THEORETICAL CALCULATIONS}

 Tight focusing due to objective lenses with a high numerical aperture (NA)  generates a non-paraxial condition. For the determination of electric and magnetic fields of radially and azimuthally polarized Lagauerre-Gaussian (LG) beams under non-paraxial conditions, we use the angular spectrum method or Vector
Diffraction theory of Richards and Wolf \cite{richards1959electromagnetic,Novotny2012}.
The electric field components (Ex, Ey, and Ez) of a focused radially polarized LG beam in the focal plane in Cartesian coordinates (x, y, and z) can be expressed as 
\onecolumngrid
\begin{multline}
\begin{split}
\begin{aligned}
\left[\begin{array}{l}
E_{x}^{o} \\
E_{y}^{o} \\
E_{z}^{o}
\end{array}\right]_{R}=A i^{m+1} \exp (i m \phi) \int_{0}^{\theta_{\max }} f_{\omega}(\theta) \cos ^{3 / 2} \theta \sin ^{2}  \theta \exp (i k z \cos \theta)  \\ ~~~~~~~~~~~~~~~~~~~~~~~\left[\begin{array}{l}
-i\left(J_{m+1} - J_{m-1}\right) \cos \phi+\left(J_{m+1}+J_{m-1}\right) \sin \phi \\
-i\left(J_{m+1} - J_{m-1}\right) \sin \phi-\left(J_{m+1}+J_{m-1}\right) \cos \phi \\
~~~~~~~~~~~~~~~~~~~2\tan \theta J_{m}
\end{array}\right] d \theta
\label{rad}
\end{aligned}
\end{split}
\end{multline}

Similarly, the electric field components of the azimuthal polarization LG beam in Cartesian coordinates (x, y, and z) can be expressed as

\begin{multline}
    \begin{aligned}
\left[\begin{array}{l}
E_{x}^{o} \\
E_{y}^{o} \\
E_{z}^{o}
\end{array}\right]_{A}=A i^{m+1} \exp (i m \phi) \int_{0}^{\theta_{\max }} f_{\omega}(\theta) \cos ^{1/ 2} \theta \sin ^{2} \theta \exp (i k z \cos \theta) \\ 
\left[\begin{array}{l}
i\left(J_{m+1}+J_{m-1}\right) \cos \phi-\left(J_{m+1}-J_{m-1}\right) \sin \phi \\
i\left(J_{m+1}+J_{m-1}\right) \sin \phi+\left(J_{m+1}-J_{m-1}\right) \cos \phi \\
~~~~~~~~~~~~~~~~~~~~~~~~~~~~~0
\end{array}\right] d \theta
\label{azi}
\end{aligned}
\end{multline}

\twocolumngrid
where, $\theta_{\max }=\sin ^{-1}(\mathrm{NA} / n)$, which is the maximum angle
related to the numerical aperture (NA) of the objective, $n$ is the refractive
index of the medium, $E^{o}$ is the output electric field, $ A $ and $B$ is the constant related to amplitude of the electric field and magnetic field respectively.
and $f_{\omega}(\theta)$ is the apodization function which appears when the beam is tightly focused by an aplanatic lens. $J_{m} $ is the $m^{th}$-order Bessel function of the first kind. $f \sin \theta_{\max }$ is the aperture radius of our lens. $\omega_{0}$ is the radius of the beam waist, $\theta$ and $\phi $ denote the tangential angle with respect to the z-axis and the azimuthal angle with respect to the x-axis, respectively. The subscripts (R) and (A) in the above equations \ref{rad} and \ref{azi} represent  the radially and azimuthally polarized beam, respectively, and $\exp (i m \phi) $ represents the helical phase, where $m$ is the topological charge that can be any integer. For non-zero values of orbital angular momentum (OAM) ($m \neq 0$), particles will revolve around the beam propagation axis \cite{garces2003observation}.

Using the above equations, we can write the expression of the electric field of radially and azimuthally polarized LG beams having zero intrinsic OAM ($m=0$)  as a linear combination of $HG_{10}$ and $HG_{01}$ mode \cite{Novotny2012, gupta2015wave}.

\begin{equation}
\left[\begin{array}{c}
E_{x}^{0} \\
E_{y}^{0} \\
E_{z}^{0}
\end{array}\right]_{Rad}=A\left[\begin{array}{c}
\left(-i I_{11}\right) \cos \phi \\
\left(-i I_{11}\right) \sin \phi \\
 I_{10}
\end{array}\right]
\end{equation}

\begin{equation}
\left[\begin{array}{c}
E_{x}^{0} \\
E_{y}^{0} \\
E_{z}^{0}
\end{array}\right]_{Azi}=A\left[\begin{array}{c}
\left(i I_{12}\right) \sin \phi \\
\left(-i I_{12}\right) \cos \phi \\
0
\end{array}\right]
\end{equation}
Here $I_{11},I_{12}$ and $ I_{10} $ are the diffraction integrals. However, when $m = 0 $, the intensity profile for the radially polarized LG beam in the focal plane appears as a bright spot at the center of the beam, since only the zero-order Bessel function $(J_{0})$ possesses a non-vanishing value at the origin. As a consequence, the longitudinal component of the electric field ($E_{z}$) arises at the focus \cite{novotny2001longitudinal}. Thus, from Eq. 1 and 2, it can be clearly seen that the z-component of the electric field emerges on tight focusing of radially polarized and azimuthally polarized LG beams for $m=0$ and $m=1$ order, respectively, while the output field for an azimuthally polarized LG beam remains purely transverse in nature at the focal plane for $m=0$.

The magnetic field for an input $m=0$ LG beam, corresponding to a radially polarized electric field, is azimuthal in nature and is given as,

\begin{equation}
\left[\begin{array}{c}
H_{x}^{0} \\
H_{y}^{0} \\
H_{z}^{0}
\end{array}\right]_{R}=B\left[\begin{array}{c}
\left(-i I_{12}\right) \cos \phi \\
\left(i I_{12}\right) \sin \phi \\
 0
\end{array}\right]
\end{equation}

\begin{equation}
\left[\begin{array}{c}
H_{x}^{0} \\
H_{y}^{0} \\
H_{z}^{0}
\end{array}\right]_{A}=B\left[\begin{array}{c}
\left(-i I_{11}\right) \sin \phi \\
\left(-i I_{11}\right) \cos \phi \\
I_{10}
\end{array}\right]
\end{equation}

where $E^{o}$  denotes the output electric field, and $I_{11}, I_{12}$ and $ I_{10} $ are the Debye–Wolf   integrals \cite{Novotny2012}. $I_{11}$ and $I_{12}$ are the integral coefficients for the transverse electric field, whereas $I_{10}$ is the coefficient for the longitudinal component of the electric field for a radially polarized zero order ($m=0$) LG beam.

$$
\begin{aligned}
&I_{11}=\int_{0}^{\theta_{\max }} f_{\omega}(\theta) \cos ^{3 / 2} \theta \sin ^{2} \theta e^{i k z \cos \theta} J_{1}(k \rho \sin \theta)  d \theta\\
&I_{10}=\int_{0}^{\theta_{m a x}} f_{\omega}(\theta) \cos ^{1/2} \theta \sin ^{3} \theta e^{i k z \cos \theta} J_{0}(k\rho \sin \theta) d \theta\\
&I_{12}=\int_{0}^{\theta_{\max }} f_{\omega}(\theta) \cos ^{1 / 2} \theta \sin ^{2} \theta e^{i k z \cos \theta} J_{1}(k \rho \sin \theta) d \theta
\end{aligned}
$$

The total intensity distribution of the output electric field for  input radially polarized LG beam is given by 

\begin{equation}
\left.I(\rho)=A^{2}\left(\mid I_{11} \right|)^{2}+\left|I_{10}\right|^{2}\right)
\end{equation}

\subsection{Numerical Simulations}

Our simulations are performed for tight focusing of the input radial/azimuthal beam by a high NA objective lens into a stratified medium as described in the main ms. The electric field in the focal plane for radial polarization exhibits a component not only along the transverse direction, but also in the longitudinal direction, since the zero-order Bessel function of the first kind $J_{0}$ is not zero at the focus of the beam. On the other hand, for azimuthal polarization, the longitudinal component of the electric field is zero. As we have mentioned previously, the electric field in the transverse plane depends on Debye–Wolf integrals $I_{11}$ for radial, $I_{12}$ for azimuthal, while the longitudinal component depends on $I_{10}$. 

\begin{center}
\begin{figure}[!h]
\includegraphics[width=0.5\textwidth]{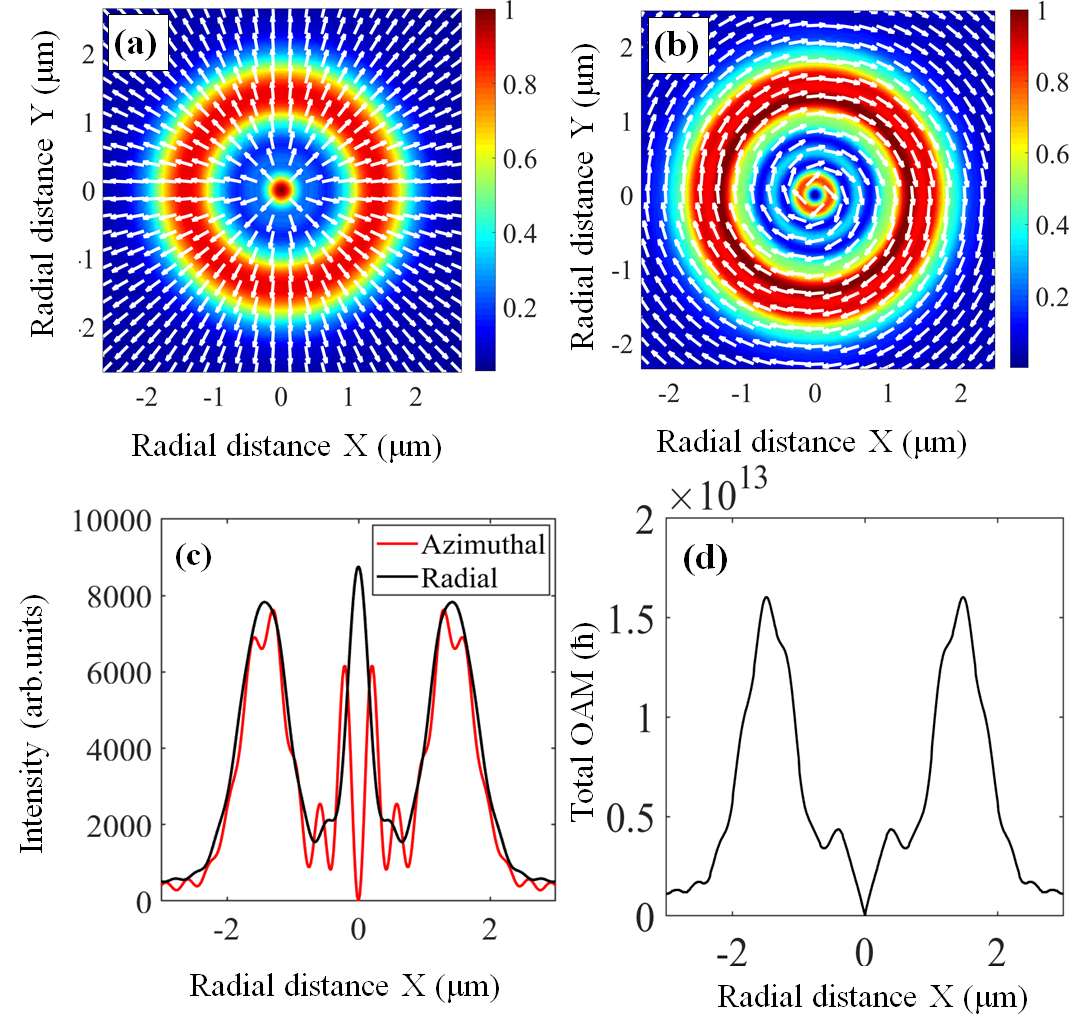}
\caption{Numerical simulation of intensity distribution at z = 2 $\mu m$
away the focus of the high NA objective (trap focus) lens for an input (a) radially polarized (b) azimuthally polarized $LG_{10}$ beam. (c) Radial distribution of intensity of radially and azimuthally polarized $LG_{10}$ beams. (d) Radial distribution of total OAM which is highly concentrated in intensity annular ring for mismatched RI (1.814) at 2 $\mu m$ away from focus.  }
\label{Intensity}
\end{figure}
\end{center}

We observe that the intensity at the center of a radially polarized beam occurs due to $I_{long}$ ( $I_{10}$), while an off-axis intensity comes from $I_{trans}$ ($I_{11}$), as shown in Fig. ~\ref{Intensity} (a) and \ref{OAM} (g). Note that the intensity at the beam center can be increased by increasing the refractive index contrast of the stratified medium. For RI 1.814, the intensity at the beam center is around 10\% more than the annular ring so that the particles are trapped in both regions. However, the intensity at the  center of an azimuthally polarized beam is found to be zero due to the absence of longitudinal component ($E_{z}$) of the electric field (Fig. ~\ref{Intensity} (b)). Therefore, particles (size $\sim$ ~ 1 $\mu m$) are only trapped in an annular ring.  In fig. ~\ref{Intensity} (c) we show the line plot of the corresponding intensity of the tightly focused radially and azimuthally polarized beams for an RI 1.814 at $2 \mu m$ away from the focus. It is clear from Fig~ \ref{Intensity} (d) that total OAM for both beams is maximum at the intensity annular ring (radial distance $\sim ~ 2 \mu m$) and zero at the beam center.

\section{Experimental methods}

We use a conventional optical tweezers configuration consisting of an inverted microscope (Carl Zeiss Axioert.A1) with an oil-immersion 100X objective (Zeiss, NA 1.4) and a solid state laser (Lasever, 671 $nm$, 350 $mW$) coupled to the back port of the microscope. We use a vortex half-wave retarder ($q$-plate) of zero-order for generating structured vector beams (i.e.radially and azimuthally polarized LG $m=0$ beam). We fix the fast axis orientation of vortex plate ($q=\frac{1}{2}$) in such a way so that it converts linear $x$- polarized and $y$-polarized light into azimuthally and radially polarized light, respectively. For the probe particles, we use RM257 vaterite particles and nematic liquid crystal which are optically anisotropic and birefringent so as to transfer angular momentum (spin and orbital) from the beam into the particles \cite{sandomirski2004highly}. This facilitates probing the effects of OAM and electric and magnetic TSAM. We then couple the radially (azimuthally) polarized LG ($m=0$) beam into the microscope so that it is tightly focused into the stratified medium described earlier. The cover slip and glass slide sandwiched together make up the sample chamber into which we add approximately $20\mu$l of the aqueous dispersion of LC and RM257 particles. The mean size of RM257 particles is $ 1-2 ~\mu$m, while that of LC droplets are $ 2-4 ~\mu$m  with a standard deviation of 20\%. The LC droplets have much higher birefringence compared to the vaterite particles, so we use them to probe effects of TSAM. The RM257 particles, on the other hand, are much smaller - so they can be trapped in the annular intensity ring, in order to probe the effects of OAM. We collect the forward-transmitted light from the microscope lamp, as well as back-reflected light from the particles for characterizing the spin and orbital rotations, respectively. The TSAM transfer to  particles trapped at the trap center, as well as the OAM transfer to particles trapped in the off-axis intensity ring, are optimized by varying the $z$-focus of the microscope objective. \par
\medskip



\begin{thebibliography}{36}%
\makeatletter
\providecommand \@ifxundefined [1]{%
 \@ifx{#1\undefined}
}%
\providecommand \@ifnum [1]{%
 \ifnum #1\expandafter \@firstoftwo
 \else \expandafter \@secondoftwo
 \fi
}%
\providecommand \@ifx [1]{%
 \ifx #1\expandafter \@firstoftwo
 \else \expandafter \@secondoftwo
 \fi
}%
\providecommand \natexlab [1]{#1}%
\providecommand \enquote  [1]{``#1''}%
\providecommand \bibnamefont  [1]{#1}%
\providecommand \bibfnamefont [1]{#1}%
\providecommand \citenamefont [1]{#1}%
\providecommand \href@noop [0]{\@secondoftwo}%
\providecommand \href [0]{\begingroup \@sanitize@url \@href}%
\providecommand \@href[1]{\@@startlink{#1}\@@href}%
\providecommand \@@href[1]{\endgroup#1\@@endlink}%
\providecommand \@sanitize@url [0]{\catcode `\\12\catcode `\$12\catcode
  `\&12\catcode `\#12\catcode `\^12\catcode `\_12\catcode `\%12\relax}%
\providecommand \@@startlink[1]{}%
\providecommand \@@endlink[0]{}%
\providecommand \url  [0]{\begingroup\@sanitize@url \@url }%
\providecommand \@url [1]{\endgroup\@href {#1}{\urlprefix }}%
\providecommand \urlprefix  [0]{URL }%
\providecommand \Eprint [0]{\href }%
\providecommand \doibase [0]{https://doi.org/}%
\providecommand \selectlanguage [0]{\@gobble}%
\providecommand \bibinfo  [0]{\@secondoftwo}%
\providecommand \bibfield  [0]{\@secondoftwo}%
\providecommand \translation [1]{[#1]}%
\providecommand \BibitemOpen [0]{}%
\providecommand \bibitemStop [0]{}%
\providecommand \bibitemNoStop [0]{.\EOS\space}%
\providecommand \EOS [0]{\spacefactor3000\relax}%
\providecommand \BibitemShut  [1]{\csname bibitem#1\endcsname}%
\let\auto@bib@innerbib\@empty
\bibitem [{\citenamefont {Yang}\ \emph {et~al.}(2021)\citenamefont {Yang},
  \citenamefont {Ren}, \citenamefont {Chen}, \citenamefont {Arita},\ and\
  \citenamefont {Rosales-Guzm{\'a}n}}]{yang2021optical}%
  \BibitemOpen
  \bibfield  {author} {\bibinfo {author} {\bibfnamefont {Y.}~\bibnamefont
  {Yang}}, \bibinfo {author} {\bibfnamefont {Y.-X.}\ \bibnamefont {Ren}},
  \bibinfo {author} {\bibfnamefont {M.}~\bibnamefont {Chen}}, \bibinfo {author}
  {\bibfnamefont {Y.}~\bibnamefont {Arita}},\ and\ \bibinfo {author}
  {\bibfnamefont {C.}~\bibnamefont {Rosales-Guzm{\'a}n}},\ }\href@noop {}
  {\bibfield  {journal} {\bibinfo  {journal} {Advanced Photonics}\ }\textbf
  {\bibinfo {volume} {3}},\ \bibinfo {pages} {034001} (\bibinfo {year}
  {2021})}\BibitemShut {NoStop}%
\bibitem [{\citenamefont {Kotlyar}\ \emph {et~al.}(2020)\citenamefont
  {Kotlyar}, \citenamefont {Nalimov}, \citenamefont {Kovalev}, \citenamefont
  {Porfirev},\ and\ \citenamefont {Stafeev}}]{kotlyar2020spin}%
  \BibitemOpen
  \bibfield  {author} {\bibinfo {author} {\bibfnamefont {V.}~\bibnamefont
  {Kotlyar}}, \bibinfo {author} {\bibfnamefont {A.}~\bibnamefont {Nalimov}},
  \bibinfo {author} {\bibfnamefont {A.}~\bibnamefont {Kovalev}}, \bibinfo
  {author} {\bibfnamefont {A.}~\bibnamefont {Porfirev}},\ and\ \bibinfo
  {author} {\bibfnamefont {S.}~\bibnamefont {Stafeev}},\ }\href@noop {}
  {\bibfield  {journal} {\bibinfo  {journal} {Physical Review A}\ }\textbf
  {\bibinfo {volume} {102}},\ \bibinfo {pages} {033502} (\bibinfo {year}
  {2020})}\BibitemShut {NoStop}%
\bibitem [{\citenamefont {Zhao}\ \emph {et~al.}(2007)\citenamefont {Zhao},
  \citenamefont {Edgar}, \citenamefont {Jeffries}, \citenamefont {McGloin},\
  and\ \citenamefont {Chiu}}]{zhao2007spin}%
  \BibitemOpen
  \bibfield  {author} {\bibinfo {author} {\bibfnamefont {Y.}~\bibnamefont
  {Zhao}}, \bibinfo {author} {\bibfnamefont {J.~S.}\ \bibnamefont {Edgar}},
  \bibinfo {author} {\bibfnamefont {G.~D.}\ \bibnamefont {Jeffries}}, \bibinfo
  {author} {\bibfnamefont {D.}~\bibnamefont {McGloin}},\ and\ \bibinfo {author}
  {\bibfnamefont {D.~T.}\ \bibnamefont {Chiu}},\ }\href@noop {} {\bibfield
  {journal} {\bibinfo  {journal} {Physical review letters}\ }\textbf {\bibinfo
  {volume} {99}},\ \bibinfo {pages} {073901} (\bibinfo {year}
  {2007})}\BibitemShut {NoStop}%
\bibitem [{\citenamefont {Friese}\ \emph {et~al.}(1998)\citenamefont {Friese},
  \citenamefont {Nieminen}, \citenamefont {Heckenberg},\ and\ \citenamefont
  {Rubinsztein-Dunlop}}]{friese1998optical}%
  \BibitemOpen
  \bibfield  {author} {\bibinfo {author} {\bibfnamefont {M.~E.}\ \bibnamefont
  {Friese}}, \bibinfo {author} {\bibfnamefont {T.~A.}\ \bibnamefont
  {Nieminen}}, \bibinfo {author} {\bibfnamefont {N.~R.}\ \bibnamefont
  {Heckenberg}},\ and\ \bibinfo {author} {\bibfnamefont {H.}~\bibnamefont
  {Rubinsztein-Dunlop}},\ }\href@noop {} {\bibfield  {journal} {\bibinfo
  {journal} {Nature}\ }\textbf {\bibinfo {volume} {394}},\ \bibinfo {pages}
  {348} (\bibinfo {year} {1998})}\BibitemShut {NoStop}%
\bibitem [{\citenamefont {Garc{\'e}s-Ch{\'a}vez}\ \emph
  {et~al.}(2003)\citenamefont {Garc{\'e}s-Ch{\'a}vez}, \citenamefont {McGloin},
  \citenamefont {Padgett}, \citenamefont {Dultz}, \citenamefont {Schmitzer},\
  and\ \citenamefont {Dholakia}}]{garces2003observation}%
  \BibitemOpen
  \bibfield  {author} {\bibinfo {author} {\bibfnamefont {V.}~\bibnamefont
  {Garc{\'e}s-Ch{\'a}vez}}, \bibinfo {author} {\bibfnamefont {D.}~\bibnamefont
  {McGloin}}, \bibinfo {author} {\bibfnamefont {M.}~\bibnamefont {Padgett}},
  \bibinfo {author} {\bibfnamefont {W.}~\bibnamefont {Dultz}}, \bibinfo
  {author} {\bibfnamefont {H.}~\bibnamefont {Schmitzer}},\ and\ \bibinfo
  {author} {\bibfnamefont {K.}~\bibnamefont {Dholakia}},\ }\href@noop {}
  {\bibfield  {journal} {\bibinfo  {journal} {Physical review letters}\
  }\textbf {\bibinfo {volume} {91}},\ \bibinfo {pages} {093602} (\bibinfo
  {year} {2003})}\BibitemShut {NoStop}%
\bibitem [{\citenamefont {Stilgoe}\ \emph {et~al.}(2022)\citenamefont
  {Stilgoe}, \citenamefont {Nieminen},\ and\ \citenamefont
  {Rubinsztein-Dunlop}}]{stilgoe2022controlled}%
  \BibitemOpen
  \bibfield  {author} {\bibinfo {author} {\bibfnamefont {A.~B.}\ \bibnamefont
  {Stilgoe}}, \bibinfo {author} {\bibfnamefont {T.~A.}\ \bibnamefont
  {Nieminen}},\ and\ \bibinfo {author} {\bibfnamefont {H.}~\bibnamefont
  {Rubinsztein-Dunlop}},\ }\href@noop {} {\bibfield  {journal} {\bibinfo
  {journal} {Nature Photonics}\ }\textbf {\bibinfo {volume} {16}},\ \bibinfo
  {pages} {346} (\bibinfo {year} {2022})}\BibitemShut {NoStop}%
\bibitem [{\citenamefont {He}\ \emph {et~al.}(1995)\citenamefont {He},
  \citenamefont {Friese}, \citenamefont {Heckenberg},\ and\ \citenamefont
  {Rubinsztein-Dunlop}}]{he1995direct}%
  \BibitemOpen
  \bibfield  {author} {\bibinfo {author} {\bibfnamefont {H.}~\bibnamefont
  {He}}, \bibinfo {author} {\bibfnamefont {M.}~\bibnamefont {Friese}}, \bibinfo
  {author} {\bibfnamefont {N.}~\bibnamefont {Heckenberg}},\ and\ \bibinfo
  {author} {\bibfnamefont {H.}~\bibnamefont {Rubinsztein-Dunlop}},\ }\href
  {https://doi.org/10.1103/PhysRevLett.75.826} {\bibfield  {journal} {\bibinfo
  {journal} {Physical review letters}\ }\textbf {\bibinfo {volume} {75}},\
  \bibinfo {pages} {826} (\bibinfo {year} {1995})}\BibitemShut {NoStop}%
\bibitem [{\citenamefont {Anh{\"a}user}\ \emph {et~al.}(2012)\citenamefont
  {Anh{\"a}user}, \citenamefont {Wunenburger},\ and\ \citenamefont
  {Brasselet}}]{anhauser2012acoustic}%
  \BibitemOpen
  \bibfield  {author} {\bibinfo {author} {\bibfnamefont {A.}~\bibnamefont
  {Anh{\"a}user}}, \bibinfo {author} {\bibfnamefont {R.}~\bibnamefont
  {Wunenburger}},\ and\ \bibinfo {author} {\bibfnamefont {E.}~\bibnamefont
  {Brasselet}},\ }\href@noop {} {\bibfield  {journal} {\bibinfo  {journal}
  {Physical review letters}\ }\textbf {\bibinfo {volume} {109}},\ \bibinfo
  {pages} {034301} (\bibinfo {year} {2012})}\BibitemShut {NoStop}%
\bibitem [{\citenamefont {Novotny}\ \emph {et~al.}(2001)\citenamefont
  {Novotny}, \citenamefont {Beversluis}, \citenamefont {Youngworth},\ and\
  \citenamefont {Brown}}]{novotny2001longitudinal}%
  \BibitemOpen
  \bibfield  {author} {\bibinfo {author} {\bibfnamefont {L.}~\bibnamefont
  {Novotny}}, \bibinfo {author} {\bibfnamefont {M.}~\bibnamefont {Beversluis}},
  \bibinfo {author} {\bibfnamefont {K.}~\bibnamefont {Youngworth}},\ and\
  \bibinfo {author} {\bibfnamefont {T.}~\bibnamefont {Brown}},\ }\href@noop {}
  {\bibfield  {journal} {\bibinfo  {journal} {Physical review letters}\
  }\textbf {\bibinfo {volume} {86}},\ \bibinfo {pages} {5251} (\bibinfo {year}
  {2001})}\BibitemShut {NoStop}%
\bibitem [{\citenamefont {Shao}\ \emph {et~al.}(2018)\citenamefont {Shao},
  \citenamefont {Zhu}, \citenamefont {Chen}, \citenamefont {Zhang},\ and\
  \citenamefont {Yu}}]{shao2018spin}%
  \BibitemOpen
  \bibfield  {author} {\bibinfo {author} {\bibfnamefont {Z.}~\bibnamefont
  {Shao}}, \bibinfo {author} {\bibfnamefont {J.}~\bibnamefont {Zhu}}, \bibinfo
  {author} {\bibfnamefont {Y.}~\bibnamefont {Chen}}, \bibinfo {author}
  {\bibfnamefont {Y.}~\bibnamefont {Zhang}},\ and\ \bibinfo {author}
  {\bibfnamefont {S.}~\bibnamefont {Yu}},\ }\href@noop {} {\bibfield  {journal}
  {\bibinfo  {journal} {Nature communications}\ }\textbf {\bibinfo {volume}
  {9}},\ \bibinfo {pages} {1} (\bibinfo {year} {2018})}\BibitemShut {NoStop}%
\bibitem [{\citenamefont {Roy}\ \emph {et~al.}(2022)\citenamefont {Roy},
  \citenamefont {Ghosh}, \citenamefont {Banerjee},\ and\ \citenamefont
  {Gupta}}]{roy2022manipulating}%
  \BibitemOpen
  \bibfield  {author} {\bibinfo {author} {\bibfnamefont {S.}~\bibnamefont
  {Roy}}, \bibinfo {author} {\bibfnamefont {N.}~\bibnamefont {Ghosh}}, \bibinfo
  {author} {\bibfnamefont {A.}~\bibnamefont {Banerjee}},\ and\ \bibinfo
  {author} {\bibfnamefont {S.~D.}\ \bibnamefont {Gupta}},\ }\href@noop {}
  {\bibfield  {journal} {\bibinfo  {journal} {Physical Review A}\ }\textbf
  {\bibinfo {volume} {105}},\ \bibinfo {pages} {063514} (\bibinfo {year}
  {2022})}\BibitemShut {NoStop}%
\bibitem [{\citenamefont {Bliokh}\ and\ \citenamefont
  {Nori}(2015)}]{bliokh2015transverse}%
  \BibitemOpen
  \bibfield  {author} {\bibinfo {author} {\bibfnamefont {K.~Y.}\ \bibnamefont
  {Bliokh}}\ and\ \bibinfo {author} {\bibfnamefont {F.}~\bibnamefont {Nori}},\
  }\href@noop {} {\bibfield  {journal} {\bibinfo  {journal} {Physics Reports}\
  }\textbf {\bibinfo {volume} {592}},\ \bibinfo {pages} {1} (\bibinfo {year}
  {2015})}\BibitemShut {NoStop}%
\bibitem [{\citenamefont {Bliokh}\ \emph {et~al.}(2013)\citenamefont {Bliokh},
  \citenamefont {Bekshaev},\ and\ \citenamefont {Nori}}]{bliokh2013dual}%
  \BibitemOpen
  \bibfield  {author} {\bibinfo {author} {\bibfnamefont {K.~Y.}\ \bibnamefont
  {Bliokh}}, \bibinfo {author} {\bibfnamefont {A.~Y.}\ \bibnamefont
  {Bekshaev}},\ and\ \bibinfo {author} {\bibfnamefont {F.}~\bibnamefont
  {Nori}},\ }\href@noop {} {\bibfield  {journal} {\bibinfo  {journal} {New
  Journal of Physics}\ }\textbf {\bibinfo {volume} {15}},\ \bibinfo {pages}
  {033026} (\bibinfo {year} {2013})}\BibitemShut {NoStop}%
\bibitem [{\citenamefont {Bekshaev}\ \emph {et~al.}(2015)\citenamefont
  {Bekshaev}, \citenamefont {Bliokh},\ and\ \citenamefont
  {Nori}}]{bekshaev2015transverse}%
  \BibitemOpen
  \bibfield  {author} {\bibinfo {author} {\bibfnamefont {A.~Y.}\ \bibnamefont
  {Bekshaev}}, \bibinfo {author} {\bibfnamefont {K.~Y.}\ \bibnamefont
  {Bliokh}},\ and\ \bibinfo {author} {\bibfnamefont {F.}~\bibnamefont {Nori}},\
  }\href {https://doi.org/10.1103/PhysRevX.5.011039} {\bibfield  {journal}
  {\bibinfo  {journal} {Physical Review X}\ }\textbf {\bibinfo {volume} {5}},\
  \bibinfo {pages} {011039} (\bibinfo {year} {2015})}\BibitemShut {NoStop}%
\bibitem [{\citenamefont {Aiello}\ \emph {et~al.}(2015)\citenamefont {Aiello},
  \citenamefont {Banzer}, \citenamefont {Neugebauer},\ and\ \citenamefont
  {Leuchs}}]{aiello2015transverse}%
  \BibitemOpen
  \bibfield  {author} {\bibinfo {author} {\bibfnamefont {A.}~\bibnamefont
  {Aiello}}, \bibinfo {author} {\bibfnamefont {P.}~\bibnamefont {Banzer}},
  \bibinfo {author} {\bibfnamefont {M.}~\bibnamefont {Neugebauer}},\ and\
  \bibinfo {author} {\bibfnamefont {G.}~\bibnamefont {Leuchs}},\ }\href@noop {}
  {\bibfield  {journal} {\bibinfo  {journal} {Nature Photonics}\ }\textbf
  {\bibinfo {volume} {9}},\ \bibinfo {pages} {789} (\bibinfo {year}
  {2015})}\BibitemShut {NoStop}%
\bibitem [{\citenamefont {Tkachenko}\ \emph {et~al.}(2020)\citenamefont
  {Tkachenko}, \citenamefont {Toftul}, \citenamefont {Esporlas}, \citenamefont
  {Maimaiti}, \citenamefont {Le~Kien}, \citenamefont {Truong},\ and\
  \citenamefont {Chormaic}}]{tkachenko2020light}%
  \BibitemOpen
  \bibfield  {author} {\bibinfo {author} {\bibfnamefont {G.}~\bibnamefont
  {Tkachenko}}, \bibinfo {author} {\bibfnamefont {I.}~\bibnamefont {Toftul}},
  \bibinfo {author} {\bibfnamefont {C.}~\bibnamefont {Esporlas}}, \bibinfo
  {author} {\bibfnamefont {A.}~\bibnamefont {Maimaiti}}, \bibinfo {author}
  {\bibfnamefont {F.}~\bibnamefont {Le~Kien}}, \bibinfo {author} {\bibfnamefont
  {V.~G.}\ \bibnamefont {Truong}},\ and\ \bibinfo {author} {\bibfnamefont
  {S.~N.}\ \bibnamefont {Chormaic}},\ }\href@noop {} {\bibfield  {journal}
  {\bibinfo  {journal} {Optica}\ }\textbf {\bibinfo {volume} {7}},\ \bibinfo
  {pages} {59} (\bibinfo {year} {2020})}\BibitemShut {NoStop}%
\bibitem [{\citenamefont {Bliokh}\ \emph {et~al.}(2014)\citenamefont {Bliokh},
  \citenamefont {Bekshaev},\ and\ \citenamefont
  {Nori}}]{bliokh2014extraordinary}%
  \BibitemOpen
  \bibfield  {author} {\bibinfo {author} {\bibfnamefont {K.~Y.}\ \bibnamefont
  {Bliokh}}, \bibinfo {author} {\bibfnamefont {A.~Y.}\ \bibnamefont
  {Bekshaev}},\ and\ \bibinfo {author} {\bibfnamefont {F.}~\bibnamefont
  {Nori}},\ }\href@noop {} {\bibfield  {journal} {\bibinfo  {journal} {Nature
  communications}\ }\textbf {\bibinfo {volume} {5}},\ \bibinfo {pages} {3300}
  (\bibinfo {year} {2014})}\BibitemShut {NoStop}%
\bibitem [{\citenamefont {Li}\ \emph {et~al.}(2017)\citenamefont {Li},
  \citenamefont {Yan}, \citenamefont {Liang}, \citenamefont {Zhang},\ and\
  \citenamefont {Yao}}]{li2017transverse}%
  \BibitemOpen
  \bibfield  {author} {\bibinfo {author} {\bibfnamefont {M.}~\bibnamefont
  {Li}}, \bibinfo {author} {\bibfnamefont {S.}~\bibnamefont {Yan}}, \bibinfo
  {author} {\bibfnamefont {Y.}~\bibnamefont {Liang}}, \bibinfo {author}
  {\bibfnamefont {P.}~\bibnamefont {Zhang}},\ and\ \bibinfo {author}
  {\bibfnamefont {B.}~\bibnamefont {Yao}},\ }\href
  {https://doi.org/10.1103/PhysRevA.95.053802} {\bibfield  {journal} {\bibinfo
  {journal} {Physical Review A}\ }\textbf {\bibinfo {volume} {95}},\ \bibinfo
  {pages} {053802} (\bibinfo {year} {2017})}\BibitemShut {NoStop}%
\bibitem [{\citenamefont {Zhan}(2009)}]{zhan2009cylindrical}%
  \BibitemOpen
  \bibfield  {author} {\bibinfo {author} {\bibfnamefont {Q.}~\bibnamefont
  {Zhan}},\ }\href {https://doi.org/10.1364/AOP.1.000001} {\bibfield  {journal}
  {\bibinfo  {journal} {Advances in Optics and Photonics}\ }\textbf {\bibinfo
  {volume} {1}},\ \bibinfo {pages} {1} (\bibinfo {year} {2009})}\BibitemShut
  {NoStop}%
\bibitem [{\citenamefont {Dorn}\ \emph {et~al.}(2003)\citenamefont {Dorn},
  \citenamefont {Quabis},\ and\ \citenamefont {Leuchs}}]{dorn2003sharper}%
  \BibitemOpen
  \bibfield  {author} {\bibinfo {author} {\bibfnamefont {R.}~\bibnamefont
  {Dorn}}, \bibinfo {author} {\bibfnamefont {S.}~\bibnamefont {Quabis}},\ and\
  \bibinfo {author} {\bibfnamefont {G.}~\bibnamefont {Leuchs}},\ }\href
  {https://doi.org/10.1103/PhysRevLett.91.233901} {\bibfield  {journal}
  {\bibinfo  {journal} {Physical review letters}\ }\textbf {\bibinfo {volume}
  {91}},\ \bibinfo {pages} {233901} (\bibinfo {year} {2003})}\BibitemShut
  {NoStop}%
\bibitem [{\citenamefont {Sato}\ and\ \citenamefont
  {Kozawa}(2009)}]{sato2009hollow}%
  \BibitemOpen
  \bibfield  {author} {\bibinfo {author} {\bibfnamefont {S.}~\bibnamefont
  {Sato}}\ and\ \bibinfo {author} {\bibfnamefont {Y.}~\bibnamefont {Kozawa}},\
  }\href {https://doi.org/10.1364/JOSAA.26.000142} {\bibfield  {journal}
  {\bibinfo  {journal} {JOSA A}\ }\textbf {\bibinfo {volume} {26}},\ \bibinfo
  {pages} {142} (\bibinfo {year} {2009})}\BibitemShut {NoStop}%
\bibitem [{\citenamefont {Liu}\ \emph {et~al.}(2019)\citenamefont {Liu},
  \citenamefont {Li}, \citenamefont {Zhang},\ and\ \citenamefont
  {Dirbeba}}]{liu2019separation}%
  \BibitemOpen
  \bibfield  {author} {\bibinfo {author} {\bibfnamefont {X.}~\bibnamefont
  {Liu}}, \bibinfo {author} {\bibfnamefont {J.}~\bibnamefont {Li}}, \bibinfo
  {author} {\bibfnamefont {Q.}~\bibnamefont {Zhang}},\ and\ \bibinfo {author}
  {\bibfnamefont {M.~G.}\ \bibnamefont {Dirbeba}},\ }\href
  {https://doi.org/10.1039/C9CP02101A} {\bibfield  {journal} {\bibinfo
  {journal} {Physical Chemistry Chemical Physics}\ }\textbf {\bibinfo {volume}
  {21}},\ \bibinfo {pages} {15339} (\bibinfo {year} {2019})}\BibitemShut
  {NoStop}%
\bibitem [{\citenamefont {Huang}\ \emph {et~al.}(2011)\citenamefont {Huang},
  \citenamefont {Shi}, \citenamefont {Cao}, \citenamefont {Li}, \citenamefont
  {Zhang},\ and\ \citenamefont {Li}}]{huang2011vector}%
  \BibitemOpen
  \bibfield  {author} {\bibinfo {author} {\bibfnamefont {K.}~\bibnamefont
  {Huang}}, \bibinfo {author} {\bibfnamefont {P.}~\bibnamefont {Shi}}, \bibinfo
  {author} {\bibfnamefont {G.}~\bibnamefont {Cao}}, \bibinfo {author}
  {\bibfnamefont {K.}~\bibnamefont {Li}}, \bibinfo {author} {\bibfnamefont
  {X.}~\bibnamefont {Zhang}},\ and\ \bibinfo {author} {\bibfnamefont
  {Y.}~\bibnamefont {Li}},\ }\href {https://doi.org/10.1364/OL.36.000888}
  {\bibfield  {journal} {\bibinfo  {journal} {Optics Letters}\ }\textbf
  {\bibinfo {volume} {36}},\ \bibinfo {pages} {888} (\bibinfo {year}
  {2011})}\BibitemShut {NoStop}%
\bibitem [{\citenamefont {Berry}(2009)}]{berry2009optical}%
  \BibitemOpen
  \bibfield  {author} {\bibinfo {author} {\bibfnamefont {M.~V.}\ \bibnamefont
  {Berry}},\ }\href@noop {} {\bibfield  {journal} {\bibinfo  {journal} {Journal
  of Optics A: Pure and Applied Optics}\ }\textbf {\bibinfo {volume} {11}},\
  \bibinfo {pages} {094001} (\bibinfo {year} {2009})}\BibitemShut {NoStop}%
\bibitem [{\citenamefont {Roy}\ \emph {et~al.}(2013)\citenamefont {Roy},
  \citenamefont {Ghosh}, \citenamefont {Gupta}, \citenamefont {Panigrahi},
  \citenamefont {Roy},\ and\ \citenamefont {Banerjee}}]{roy2013controlled}%
  \BibitemOpen
  \bibfield  {author} {\bibinfo {author} {\bibfnamefont {B.}~\bibnamefont
  {Roy}}, \bibinfo {author} {\bibfnamefont {N.}~\bibnamefont {Ghosh}}, \bibinfo
  {author} {\bibfnamefont {S.~D.}\ \bibnamefont {Gupta}}, \bibinfo {author}
  {\bibfnamefont {P.~K.}\ \bibnamefont {Panigrahi}}, \bibinfo {author}
  {\bibfnamefont {S.}~\bibnamefont {Roy}},\ and\ \bibinfo {author}
  {\bibfnamefont {A.}~\bibnamefont {Banerjee}},\ }\href
  {https://doi.org/10.1103/PhysRevA.87.043823} {\bibfield  {journal} {\bibinfo
  {journal} {Physical Review A}\ }\textbf {\bibinfo {volume} {87}},\ \bibinfo
  {pages} {043823} (\bibinfo {year} {2013})}\BibitemShut {NoStop}%
\bibitem [{\citenamefont {Roy}\ \emph {et~al.}(2014)\citenamefont {Roy},
  \citenamefont {Ghosh}, \citenamefont {Banerjee}, \citenamefont {Gupta},\ and\
  \citenamefont {Roy}}]{roy2014manifestations}%
  \BibitemOpen
  \bibfield  {author} {\bibinfo {author} {\bibfnamefont {B.}~\bibnamefont
  {Roy}}, \bibinfo {author} {\bibfnamefont {N.}~\bibnamefont {Ghosh}}, \bibinfo
  {author} {\bibfnamefont {A.}~\bibnamefont {Banerjee}}, \bibinfo {author}
  {\bibfnamefont {S.~D.}\ \bibnamefont {Gupta}},\ and\ \bibinfo {author}
  {\bibfnamefont {S.}~\bibnamefont {Roy}},\ }\href@noop {} {\bibfield
  {journal} {\bibinfo  {journal} {New Journal of Physics}\ }\textbf {\bibinfo
  {volume} {16}},\ \bibinfo {pages} {083037} (\bibinfo {year}
  {2014})}\BibitemShut {NoStop}%
\bibitem [{\citenamefont {Pal}\ \emph {et~al.}(2020)\citenamefont {Pal},
  \citenamefont {Gupta}, \citenamefont {Ghosh},\ and\ \citenamefont
  {Banerjee}}]{pal2020direct}%
  \BibitemOpen
  \bibfield  {author} {\bibinfo {author} {\bibfnamefont {D.}~\bibnamefont
  {Pal}}, \bibinfo {author} {\bibfnamefont {S.~D.}\ \bibnamefont {Gupta}},
  \bibinfo {author} {\bibfnamefont {N.}~\bibnamefont {Ghosh}},\ and\ \bibinfo
  {author} {\bibfnamefont {A.}~\bibnamefont {Banerjee}},\ }\href
  {https://doi.org/10.1063/5.0015991} {\bibfield  {journal} {\bibinfo
  {journal} {APL Photonics}\ }\textbf {\bibinfo {volume} {5}},\ \bibinfo
  {pages} {086106} (\bibinfo {year} {2020})}\BibitemShut {NoStop}%
\bibitem [{\citenamefont {Youngworth}\ and\ \citenamefont
  {Brown}(2000)}]{youngworth2000focusing}%
  \BibitemOpen
  \bibfield  {author} {\bibinfo {author} {\bibfnamefont {K.~S.}\ \bibnamefont
  {Youngworth}}\ and\ \bibinfo {author} {\bibfnamefont {T.~G.}\ \bibnamefont
  {Brown}},\ }\href {https://doi.org/10.1364/OE.7.000077} {\bibfield  {journal}
  {\bibinfo  {journal} {Optics Express}\ }\textbf {\bibinfo {volume} {7}},\
  \bibinfo {pages} {77} (\bibinfo {year} {2000})}\BibitemShut {NoStop}%
\bibitem [{\citenamefont {O'neil}\ \emph {et~al.}(2002)\citenamefont {O'neil},
  \citenamefont {MacVicar}, \citenamefont {Allen},\ and\ \citenamefont
  {Padgett}}]{o2002intrinsic}%
  \BibitemOpen
  \bibfield  {author} {\bibinfo {author} {\bibfnamefont {A.}~\bibnamefont
  {O'neil}}, \bibinfo {author} {\bibfnamefont {I.}~\bibnamefont {MacVicar}},
  \bibinfo {author} {\bibfnamefont {L.}~\bibnamefont {Allen}},\ and\ \bibinfo
  {author} {\bibfnamefont {M.}~\bibnamefont {Padgett}},\ }\href
  {https://doi.org/10.1103/PhysRevLett.88.053601} {\bibfield  {journal}
  {\bibinfo  {journal} {Physical review letters}\ }\textbf {\bibinfo {volume}
  {88}},\ \bibinfo {pages} {053601} (\bibinfo {year} {2002})}\BibitemShut
  {NoStop}%
\bibitem [{\citenamefont {Kumar}\ \emph {et~al.}(2022)\citenamefont {Kumar},
  \citenamefont {Gupta}, \citenamefont {Ghosh}, \citenamefont {Banerjee} \emph
  {et~al.}}]{kumar2022probing}%
  \BibitemOpen
  \bibfield  {author} {\bibinfo {author} {\bibfnamefont {R.~N.}\ \bibnamefont
  {Kumar}}, \bibinfo {author} {\bibfnamefont {S.~D.}\ \bibnamefont {Gupta}},
  \bibinfo {author} {\bibfnamefont {N.}~\bibnamefont {Ghosh}}, \bibinfo
  {author} {\bibfnamefont {A.}~\bibnamefont {Banerjee}}, \emph {et~al.},\
  }\href@noop {} {\bibfield  {journal} {\bibinfo  {journal} {Physical Review
  A}\ }\textbf {\bibinfo {volume} {105}},\ \bibinfo {pages} {023503} (\bibinfo
  {year} {2022})}\BibitemShut {NoStop}%
\bibitem [{\citenamefont {Neugebauer}\ \emph {et~al.}(2015)\citenamefont
  {Neugebauer}, \citenamefont {Bauer}, \citenamefont {Aiello},\ and\
  \citenamefont {Banzer}}]{neugebauer2015measuring}%
  \BibitemOpen
  \bibfield  {author} {\bibinfo {author} {\bibfnamefont {M.}~\bibnamefont
  {Neugebauer}}, \bibinfo {author} {\bibfnamefont {T.}~\bibnamefont {Bauer}},
  \bibinfo {author} {\bibfnamefont {A.}~\bibnamefont {Aiello}},\ and\ \bibinfo
  {author} {\bibfnamefont {P.}~\bibnamefont {Banzer}},\ }\href@noop {}
  {\bibfield  {journal} {\bibinfo  {journal} {Physical review letters}\
  }\textbf {\bibinfo {volume} {114}},\ \bibinfo {pages} {063901} (\bibinfo
  {year} {2015})}\BibitemShut {NoStop}%
\bibitem [{\citenamefont {Vaippully}\ \emph {et~al.}(2021)\citenamefont
  {Vaippully}, \citenamefont {Lokesh},\ and\ \citenamefont
  {Roy}}]{vaippully2021continuous}%
  \BibitemOpen
  \bibfield  {author} {\bibinfo {author} {\bibfnamefont {R.}~\bibnamefont
  {Vaippully}}, \bibinfo {author} {\bibfnamefont {M.}~\bibnamefont {Lokesh}},\
  and\ \bibinfo {author} {\bibfnamefont {B.}~\bibnamefont {Roy}},\ }\href@noop
  {} {\bibfield  {journal} {\bibinfo  {journal} {Journal of Optics}\ }\textbf
  {\bibinfo {volume} {23}},\ \bibinfo {pages} {094001} (\bibinfo {year}
  {2021})}\BibitemShut {NoStop}%
\bibitem [{\citenamefont {Richards}\ and\ \citenamefont
  {Wolf}(1959)}]{richards1959electromagnetic}%
  \BibitemOpen
  \bibfield  {author} {\bibinfo {author} {\bibfnamefont {B.}~\bibnamefont
  {Richards}}\ and\ \bibinfo {author} {\bibfnamefont {E.}~\bibnamefont
  {Wolf}},\ }\href {https://doi.org/10.1098/rspa.1959.0200} {\bibfield
  {journal} {\bibinfo  {journal} {Proceedings of the Royal Society of London.
  Series A. Mathematical and Physical Sciences}\ }\textbf {\bibinfo {volume}
  {253}},\ \bibinfo {pages} {358} (\bibinfo {year} {1959})}\BibitemShut
  {NoStop}%
\bibitem [{\citenamefont {Novotny}\ and\ \citenamefont
  {Hecht}(2012)}]{Novotny2012}%
  \BibitemOpen
  \bibfield  {author} {\bibinfo {author} {\bibfnamefont {L.}~\bibnamefont
  {Novotny}}\ and\ \bibinfo {author} {\bibfnamefont {B.}~\bibnamefont
  {Hecht}},\ }\href@noop {} {\emph {\bibinfo {title} {Principles of
  Nano-optics}}}\ (\bibinfo  {publisher} {Cambridge university press},\
  \bibinfo {year} {2012})\BibitemShut {NoStop}%
\bibitem [{\citenamefont {Gupta}\ \emph {et~al.}(2015)\citenamefont {Gupta},
  \citenamefont {Ghosh},\ and\ \citenamefont {Banerjee}}]{gupta2015wave}%
  \BibitemOpen
  \bibfield  {author} {\bibinfo {author} {\bibfnamefont {S.~D.}\ \bibnamefont
  {Gupta}}, \bibinfo {author} {\bibfnamefont {N.}~\bibnamefont {Ghosh}},\ and\
  \bibinfo {author} {\bibfnamefont {A.}~\bibnamefont {Banerjee}},\ }\href@noop
  {} {\emph {\bibinfo {title} {Wave optics: Basic concepts and contemporary
  trends}}}\ (\bibinfo  {publisher} {CRC Press},\ \bibinfo {year}
  {2015})\BibitemShut {NoStop}%
\bibitem [{\citenamefont {Sandomirski}\ \emph {et~al.}(2004)\citenamefont
  {Sandomirski}, \citenamefont {Martin}, \citenamefont {Maret}, \citenamefont
  {Stark},\ and\ \citenamefont {Gisler}}]{sandomirski2004highly}%
  \BibitemOpen
  \bibfield  {author} {\bibinfo {author} {\bibfnamefont {K.}~\bibnamefont
  {Sandomirski}}, \bibinfo {author} {\bibfnamefont {S.}~\bibnamefont {Martin}},
  \bibinfo {author} {\bibfnamefont {G.}~\bibnamefont {Maret}}, \bibinfo
  {author} {\bibfnamefont {H.}~\bibnamefont {Stark}},\ and\ \bibinfo {author}
  {\bibfnamefont {T.}~\bibnamefont {Gisler}},\ }\href@noop {} {\bibfield
  {journal} {\bibinfo  {journal} {Journal of Physics: Condensed Matter}\
  }\textbf {\bibinfo {volume} {16}},\ \bibinfo {pages} {S4137} (\bibinfo {year}
  {2004})}\BibitemShut {NoStop}%
\end{thebibliography}
\providecommand{\noopsort}[1]{}\providecommand{\singleletter}[1]{#1}%

\end{document}